\documentclass[12pt,twoside]{article} 
\usepackage{graphicx}
\usepackage{epsfig}
\usepackage{amssymb}
\usepackage{cite}

\date{}

\begin{document}

\centerline{\bf Applied Mathematical Sciences, Vol. 8, 2014, no. 37, 1837 - 1845} 

\centerline{\bf HIKARI Ltd, \ www.m-hikari.com}

\centerline{\bf http://dx.doi.org/10.12988/ams.2014.4288}

\centerline{} 

\centerline{} 

\centerline {\Large{\bf The Friendship Paradox in Scale-Free Networks}} 

\centerline{} 

\centerline{\bf {Marcos Amaku, Rafael I. Cipullo, Jos\'{e} H. H. Grisi-Filho,}}
\centerline{\bf {Fernando S. Marques and Raul Ossada}} 

\centerline{} 

\centerline{Faculdade de Medicina Veterin\'{a}ria e Zootecnia} 

\centerline{Universidade de S\~{a}o Paulo} 

\centerline{S\~{a}o Paulo, SP, 05508-270, Brazil} 


\centerline{}

\bigskip

{\footnotesize Copyright $\copyright$ 2014 Marcos Amaku et al. This is an open access article distributed under the Creative Commons Attribution License, which
permits unrestricted use, distribution, and reproduction in any medium, provided the original work is properly cited.}

\begin{abstract}
	Our friends have more friends than we do. That is the basis of the friendship paradox. 
In mathematical terms, the mean number of friends of friends is higher than the mean number of friends. 
In the present study, we analyzed the relationship between the mean degree of vertices (individuals), $\langle k \rangle$, 
and the mean number of friends of friends, $\langle k_{FF} \rangle$, in scale-free networks with degrees ranging 
from a minimum degree ($k_{min}$) to a maximum degree ($k_{max}$). We deduced an expression 
for $\langle k_{FF} \rangle - \langle k \rangle$ for scale-free networks following 
a power-law distribution with a given scaling parameter ($\alpha$). 
Based on this expression, we can quantify how the degree distribution of a scale-free network affects the mean number of friends of friends.
\end{abstract}	


\section{Introduction}
On average, our friends have more friends than we do. This statement is the basis of the friendship paradox, 
which was termed the class-size paradox by Feld \cite{Feld}. The friendship paradox is based on an interesting
property of human social networks: on average, the friends of randomly selected individuals are more central
(regarding centrality measures) than those randomly selected people \cite{Feld,Zuckerman,Newman2003,Christakis}.  
The friendship paradox offers a strategic path around the problem of sampling a social network and detecting
an emerging outbreak~\cite{Wilson2010}.

Considering $n$ individuals, 
each of which possess $k_{i}$ friends, the mean number of friends $\langle k \rangle$ is given by:
\begin{equation}
\langle k \rangle = \frac{\sum_{i=1}^{n} k_{i}}{n} \, .
\label{eq:1}
\end{equation} 
 						
	In network terminology~\cite{Caldarelli,Newman2010}, individuals are the vertices of the network, 
and a friendship between two persons is an edge. The number of edges per vertex is the degree of the vertex. 
Thus, $\langle k \rangle$ is the mean degree of the vertices.

	The total number of friends of friends is $\sum_{i=1}^{n} k_{i}^{2}$ \cite{Feld} because, 
if an individual has $k_{i}$ friends, $k_{i}$ friends contribute 
to the final sum $k_{i}$ times, resulting in a contribution of $k_{i}^{2}$ for that individual.
In Appendix A, we show that the total number of friends of friends is $\sum_{i=1}^{n} k_{i}^{2}$.
	
	The total number of friends of individuals is $\sum_{i=1}^{n} k_{i}$; thus, the mean number of friends of friends 
can be expressed as:
\begin{equation}
\langle k_{FF} \rangle = \frac{\sum k_{i}^{2}}{\sum k_{i}} \, .
\label{eq:2}
\end{equation}
 				
Considering that $\langle k^{2} \rangle = \frac{\sum_{i=1}^{n} k_{i}^{2}}{n}$ and
$\langle k \rangle = \frac{\sum_{i=1}^{n} k_{i}}{n}$, we obtain
\begin{equation}
\langle k_{FF} \rangle = \frac{\langle k^{2} \rangle}{\langle k \rangle} \, .
\label{eq:3}
\end{equation}  				
 					
The variance ($\sigma^{2}$) of the number of friends is 
\begin{equation}
\sigma^{2} = \langle k^{2} \rangle - \langle k \rangle ^{2} 
\label{eq:4}
\end{equation}

\noindent and, dividing it by $ \langle k \rangle $, we may write
\begin{equation}
\frac{\sigma^{2}}{\langle k \rangle} = \frac{\langle k^{2} \rangle}{\langle k \rangle}  - \langle k \rangle \, .
\label{eq:5}
\end{equation}		

Therefore, based on equations (\ref{eq:3}) and (\ref{eq:5}), the mean number of friends of friends is:
\begin{equation}
\langle k_{FF} \rangle = \langle k \rangle + \frac{\sigma^{2}}{\langle k \rangle} \, ,
\label{eq:6}
\end{equation}	
which is also the result obtained by Feld~\cite{Feld}. Hence, as the ratio between the variance and the mean increases, 
the difference between the mean number of friends of individuals and the mean number of friends of friends also increases.

\smallskip
{\it {\bf Remark.} We may derive the equation for $\langle k_{FF} \rangle$ taking into account the degree distribution, $P(k)$.
Assuming that edges are formed at random, the probability that a friend of a person has degree $k$ is given approximately by
\begin{equation}
\frac{k P(k)}{\sum_{k} k P(k)} = \frac{k P(k)}{\langle k \rangle} \, .
\label{eq:R1}
\end{equation}

Thus, the average degree of friends is
\begin{equation}
\langle k_{FF} \rangle = \sum_{k} k \frac{k P(k)}{\langle k \rangle} = \frac{\langle k^{2} \rangle}{\langle k \rangle} \, .
\label{eq:R2}
\end{equation} 
\noindent
}

Scale-free networks follow a power-law degree distribution:
\begin{equation}
P(k) = C k^{-\alpha} \, ,
\label{eq:Pk}
\end{equation}
where $\alpha$ is a scaling parameter and $C$ is a normalization constant. 
Many real-world networks \cite{Amaral,Caldarelli,Clauset} are scale-free. For instance, a power-law
distribution of the number of sexual partners was observed in a network of
human sexual contacts \cite{Liljeros}. As mentioned in \cite{Liljeros},
such finding may have epidemiological implications, as epidemics arise and propagate
faster in scale-free networks than in single-scale networks.

Equation (\ref{eq:Pk}) is applicable for $\alpha > 1$. A remarkable characteristic 
of scale-free networks is the presence of highly connected vertices, which are often called hubs. 
Therefore, the friendship paradox, which is a more general principle, is expected in scale-free 
networks in which the degree of hubs exceeds the average degree of the network. In the present study, 
we analyzed the relationship between the mean degree of vertices (individuals) and 
the mean number of friends of friends in scale-free networks with degrees ranging between 
a minimum degree of $k_{min}$ and a maximum degree of $k_{max}$. 

\section{The friendship paradox in scale-free networks}
For a normalized power-law distribution with degrees ranging between $k_{min}$ and $k_{max}$, we have that
\begin{equation}
\int_{k_{min}}^{k_{max}} P(k) dk = 1 \, .
\label{eq:7}
\end{equation}

Thus, the normalization constant $C = C(k_{min},k_{max})$ may be obtained from the previous equation as
\begin{equation}
C(k_{min},k_{max}) = \frac{1-\alpha}{k_{max}^{1-\alpha} - k_{min}^{1-\alpha}} \, .
\label{eq:8}
\end{equation}
In particular, when $k_{min} = 1$ and $k_{max} \rightarrow \infty$, $C = \alpha - 1$ for $\alpha > 1$.

For a scale-free network, the mean degree of vertices, 
\begin{equation}
\langle k \rangle = \int_{k_{min}}^{k_{max}} k P(k) dk \, ,
\label{eq:9}
\end{equation}
and the variance, 
\begin{equation}
\sigma^{2} = \langle k^{2} \rangle - \langle k \rangle^{2} \, ,
\label{eq:10}
\end{equation}
are, respectively, given by
\begin{equation}
\langle k \rangle = \left( \frac{\alpha - 1}{\alpha - 2}\right)
\left( \frac{k_{max}^{2-\alpha} - k_{min}^{2-\alpha}}{k_{max}^{1-\alpha} - k_{min}^{1-\alpha}}\right)
\label{eq:11}
\end{equation}
and
\begin{eqnarray}
\sigma^{2} & = & \left( \frac{\alpha - 1}{\alpha - 3}\right)
\left( \frac{k_{max}^{3-\alpha} - k_{min}^{3-\alpha}}{k_{max}^{1-\alpha} - k_{min}^{1-\alpha}}\right) \nonumber \\
& - & \left[ 
\left( \frac{\alpha - 1}{\alpha - 2}\right)
\left( \frac{k_{max}^{2-\alpha} - k_{min}^{2-\alpha}}{k_{max}^{1-\alpha} - k_{min}^{1-\alpha}}\right)
\right]^{2} \, .
\label{eq:12}
\end{eqnarray}

Note that, for $k_{max} \rightarrow \infty$ , when $1 < \alpha < 2 $, the mean and variance are infinite. 
When $2 < \alpha < 3 $, the mean is finite, but the variance is infinite. However, when $k_{max}$ is finite, 
which is the case in many real world networks that follow approximately a power-law degree distribution, 
both the mean and variance are finite. The scaling parameter usually lies in the range of $1.5 < \alpha < 3 $, 
even though there are exceptions~\cite{Clauset}.

From equations (\ref{eq:11}) and (\ref{eq:12}), we can derive the ratio between the variance and the mean
\begin{eqnarray}
\frac{\sigma^{2}}{\langle k \rangle} & = & \left( \frac{\alpha - 2}{\alpha - 3}\right)
\left( \frac{k_{max}^{3-\alpha} - k_{min}^{3-\alpha}}{k_{max}^{2-\alpha} - k_{min}^{2-\alpha}}\right) \nonumber \\
& - &
\left( \frac{\alpha - 1}{\alpha - 2}\right)
\left( \frac{k_{max}^{2-\alpha} - k_{min}^{2-\alpha}}{k_{max}^{1-\alpha} - k_{min}^{1-\alpha}}\right) \, ,
\label{eq:13}
\end{eqnarray}	 		
which is equivalent to the difference $\langle k_{FF} \rangle - \langle k \rangle$, as shown in equation (\ref{eq:6}).	

We can derive $\langle k_{FF} \rangle$ from equations (\ref{eq:11}) and (\ref{eq:13})
\begin{equation}
\langle k_{FF} \rangle = 
\left( \frac{\alpha - 2}{\alpha - 3}\right)
\left( \frac{k_{max}^{3-\alpha} - k_{min}^{3-\alpha}}{k_{max}^{2-\alpha} - k_{min}^{2-\alpha}}\right) \, .
\label{eq:14}
\end{equation}

\section{Singularities}
Both $\langle k_{FF} \rangle$ and $\langle k \rangle$ for $\alpha = 2$ 
and $\langle k_{FF} \rangle$ for $\alpha = 3$ assume the indeterminate form $0/0$. 
By applying L'H\^{o}pital's rule, the indeterminacy was removed, 
and we obtained the following expressions for $\langle k \rangle$
\begin{equation}
\lim_{\alpha \rightarrow 2} \langle k \rangle = 
\frac{k_{min} k_{max} \ln\left( \frac{k_{max}}{k_{min}} \right)}{k_{max} - k_{min}} \, ,
\label{eq:15}
\end{equation} 
for $\langle k_{FF} \rangle$
\begin{equation}
\langle k_{FF} \rangle = \left\{ 
\begin{array}{l}
\frac{k_{max} - k_{min}}{\ln\left( \frac{k_{max}}{k_{min}} \right)} \, , \quad \quad \quad \, \, \alpha \rightarrow 2  \\
\frac{k_{min} k_{max} \ln\left( \frac{k_{max}}{k_{min}} \right)}{k_{max} - k_{min}} \, , \, \alpha \rightarrow 3 \, ,
\end{array}
\right.
\label{eq:16}
\end{equation}
and the difference $\langle k_{FF} \rangle - \langle k \rangle$ 
\begin{equation}
\langle k_{FF} \rangle - \langle k \rangle = \left\{ 
\begin{array}{l}
\frac{k_{max} - k_{min}}{\ln\left( \frac{k_{max}}{k_{min}} \right)} 
- \frac{k_{min} k_{max} \ln\left( \frac{k_{max}}{k_{min}} \right)}{k_{max} - k_{min}}
, \alpha \rightarrow 2  \\
\frac{k_{min} k_{max} \ln\left( \frac{k_{max}}{k_{min}} \right)}{k_{max} - k_{min}} 
- \frac{2 k_{min} k_{max}}{k_{max} + k_{min}}
, \alpha \rightarrow 3 .
\end{array}
\right.
\label{eq:17}
\end{equation}

\section{Results and Discussion}
\label{sec:results}
\begin{figure}
\centering
\scalebox{0.6}{\includegraphics{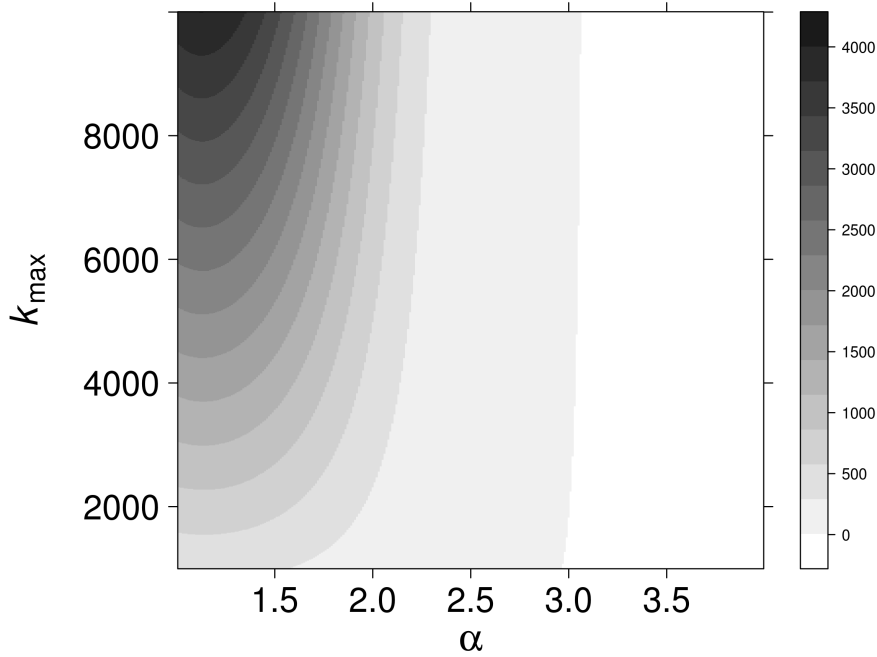}}
\caption{Variance-to-mean ratio, which is equivalent to the difference $\langle k_{FF} \rangle - \langle k \rangle$
as a function of $k_{max}$, the maximum degree observed in the network, and the scaling parameter $\alpha$, 
for a minimum degree of $k_{min}=1$.}
\label{fig.1}
\end{figure}

Based on equation (\ref{eq:13}), we analyzed the effect of varying $\alpha$ and $k_{max}$ 
on the difference between the mean of friends of friends and the mean of friends (Fig. \ref{fig.1}). 
For networks with the same value of $k_{max}$, as the scaling parameter $\alpha$ decreased, 
an increase in the variance-to-mean ratio was observed. This finding is consistent with the fact that 
networks with $\alpha$ values close to 1 are denser than networks with higher values of $\alpha$ (for instance, closer to 3).
Furthermore, the probability of finding a hub (a highly connected vertex) with a given degree of $k_{hub}$ is lower 
for a scale-free network with a higher $\alpha$ value.

\begin{figure}
\centering
\scalebox{0.5}{\includegraphics{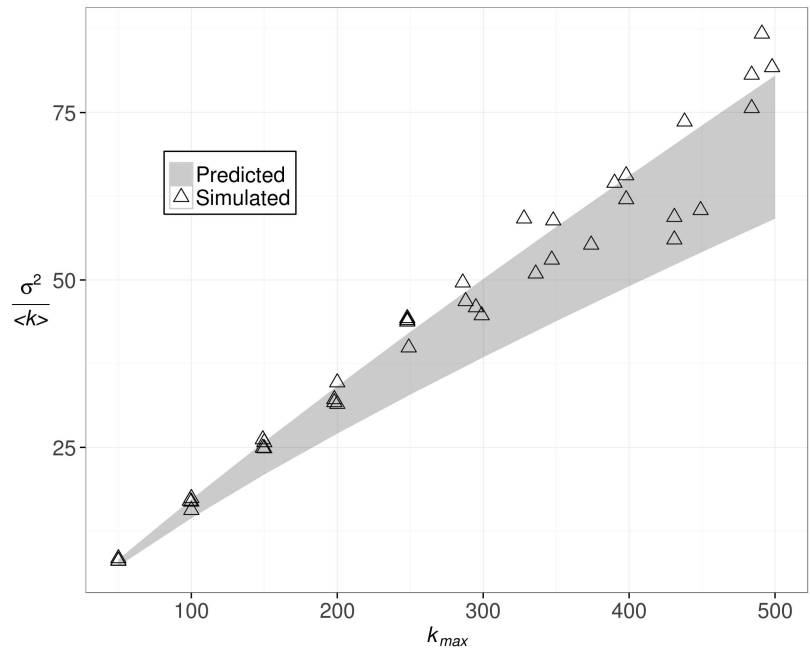}}
\caption{Variance-to-mean ratio as a function of $k_{max}$ for networks generated by simulations 
based on Models A, B and Kalisky (see~\cite{Grisi} for details) with $\alpha = 2$ and for theoretical predictions 
based on equations (\ref{eq:13}) and (\ref{eq:17}). The gray area indicates the range of predicted values 
corresponding to the fitted $\alpha$ values.}
\label{fig.2}
\end{figure}

In a previous publication, Grisi et al.~\cite{Grisi} showed that scale-free networks with the same degree distribution
may have different structures. Based on the algorithms described by Grisi et al.~\cite{Grisi}, we compared three different types of
networks with the same degree distribution and calculated the difference between the mean degree of friends of friends 
and the mean degree based on the adjacency matrix of the networks using different values of $k_{max}$. 
In Fig.~\ref{fig.2}, the simulated results of a power-law degree distribution generated using the transformation 
method~\cite{Clauset, Newman2005} with $\alpha=2$ are presented. The models used to generate the networks 
included Models A, B and Kalisky, employing the algorithms provided in \cite{Grisi}. 
In Appendix B, we summarize some characteristics of models A, B and Kalisky. 
In the simulations, we obtained similar results for the three models. A discrepancy between the simulated and predicted 
(according to the theory described in the present paper) results was observed, probably due to fluctuations 
in the generation of the degrees of the vertices --- the proportion of vertices with degree $k$ in the generated network 
may be different from the theoretical $P(k)$ ---, or due to the rounding to discrete values of continuous numbers 
generated and uncertainties in the estimation of $\alpha$. Regarding the latter factor, 
we estimated $\alpha$ using the fitting procedures described by Clauset et al.~\cite{Clauset}. 
For each value of $k_{max}$, the range of predicted values for the variance-to-mean ratio corresponding 
to the fitted $\alpha$ values are shown in gray in Fig.~\ref{fig.2}. The results illustrated in Fig.~\ref{fig.2} 
suggest that, in real networks with a given set of parameters ($\alpha$, $k_{min}$ and $k_{max}$), 
values close to the predicted results are expected, along with fluctuations.

The difference $\langle k_{FF} \rangle - \langle k \rangle$ is strongly dependent on the scaling parameter ($\alpha$) of the power-law degree distribution. 
In scale-free networks with lower $\alpha$ values, this difference is higher, reflecting the fact that the hubs in these networks are more connected
than the other vertices, in comparison to what happens in networks with higher $\alpha$.
Additionally, we would expect that information (rumors, viruses, gossip and news, among others) would spread more rapidly 
in the dense scale-free networks with lower $\alpha$ values.




	As noted by Clauset et al.~\cite{Clauset}, the characterization of power-laws is complicated by large fluctuations in the tail of the distribution. 
Provided that a dataset is derived from a degree distribution that follows a power-law in the range between $k_{min}$ and $k_{max}$, 
the expressions deduced in the present study can be used to estimate the scaling parameter $\alpha$ using equations (\ref{eq:11}), (\ref{eq:12}) or (\ref{eq:13}) 
for a given combination of $k_{min}$, $k_{max}$ and either the mean, variance, or variance-to-mean ratio.

	In summary, we deduced an expression for the difference $\langle k_{FF} \rangle - \langle k \rangle$ 
for scale-free networks that possess a maximum degree of $k_{max}$ 
and follow a power-law distribution with a scaling parameter of $\alpha$. 
Based on this expression, we can quantify how the degree distribution 
of a scale-free network affects the mean number of friends of friends. 
The intensity of $\langle k_{FF} \rangle - \langle k \rangle$, 
which increases with a decrease in the scaling parameter, 
directly affects the effectiveness of strategies for the control of infectious diseases 
(such as the strategies proposed by Christakis and Fowler~\cite{Christakis} and Cohen et al.~\cite{Cohen}), rumors or computer viruses in real scale-free networks. 
The calculations given here are also, to some extent, relevant to sampling procedures 
for monitoring and surveillance purposes in networks of human contacts and networks of animal movements.

\bigskip
{\bf Appendix A. The total number of friends of friends}

An individual $j$ has $k_{j} = \sum_{i} a_{ij}$ friends. Considering an undirected network, the 
number of friends of friends of individual $i$ is
\begin{equation}
f_{i} = \sum_{j} a_{ij} k_{j} \, .
\end{equation}

Thus, the total number of friends of friends, taking into account all individuals, is
\begin{equation}
F = \sum_{i} f_{i} = \sum_{i} \sum_{j} a_{ij} k_{j} = \sum_{j} k_{j} \sum_{i} a_{ij} = \sum_{j} k_{j}^{2} \, .
\end{equation}

\bigskip
{\bf Appendix B. Network models}

As mentioned in Section~\ref{sec:results}, models A, B and Kalisky ~\cite{Grisi, Ossada}
were used to generate the networks. The algorithms are described in Grisi et al.~\cite{Grisi}.
In this appendix, we summarize some characteristics of these models.

The networks generated by models A and Kalisky show a medium to high global efficiency, which
quantifies the efficiency of the network in sending information between vertices~\cite{Grisi}
and also a medium to high central point dominance, a measure related to the betweenness
centrality of the most central vertex in a network~\cite{Grisi}. For denser networks (see the examples in~\cite{Grisi}),
model A and Kalisky generate networks with almost all vertices in the giant component~\cite{Grisi}.

Model B, on the other hand, generates networks with very low to low global efficiency and very low
to low central point dominance. Even for denser networks (see~\cite{Grisi}),
model B generates networks with several components.
In simulations for the spread of infectious diseases~\cite{Ossada},
compared with networks generated by Model A and Kalisky
among other algorithms, the lowest prevalences of disease were observed in Model B networks.
The distribution of links in a Model B network is a plausible cause for the low prevalence, because
a large number of vertices are not connected to the giant component of the network.

\section*{Acknowledgments}
This work was partially supported by Fapesp, Capes and CNPq.
We thank an anonymous researcher for suggesting
the alternative way of deriving the average degree of friends presented in the remark.

%
%
%
%

{\bf Received: February 7, 2014}


\begin{thebibliography}{99}

\bibitem{Feld} 
{S.L. Feld, Why your friends have more friends than you do, 
\em American Journal of Sociology, } 
{\bf 96(6)} (1991), 1464 - 1477.

\bibitem{Zuckerman}
{E.W. Zuckerman and J.T. Jost,
What makes you think you're so popular? Self-evaluation maintenance and
the subjective side of the ``friendship paradox'',
\em Social Psychology Quarterly,} 
{\bf 64(3)} (2001), 207 - 223.

\bibitem{Newman2003}
{M.E.J. Newman,
Ego-centered networks and the ripple effect,
\em Social Networks,} 
{\bf 25} (2003,) 83 - 95.

\bibitem{Christakis}
{N.A. Christakis and J.H. Fowler,
Social network sensors for early detection of contagious outbreaks,
\em PLoS One,} 
{\bf 5} (2010), e12948.

\bibitem{Wilson2010}
{M. Wilson,
Using the friendship paradox to sample a social network,
\em Physics Today,}
{\bf 63(11)} (2010), 15 - 16.

\bibitem{Caldarelli}
{G. Caldarelli, \em Scale-free Networks,}
Oxford U.P., Oxford, 2007.

\bibitem{Newman2010}
{M.E.J. Newman, 
\em Networks: An Introduction,}
Oxford U.P., Oxford, 2010.

\bibitem{Amaral} 
{L.A.N. Amaral, A. Scala, M. Barth\'{e}l\'{e}my and H. E. Stanley,
Classes of small-world networks,
\em Proceedings of the National Academy of Sciences of the USA,} 
{\bf 97} (2000), 11149.

\bibitem{Clauset}
{A. Clauset, C.R. Shalizi and M.E.J. Newman,
Power-law distributions in empirical data,
\em SIAM Review,}
{\bf 51(4)} 2009, 661 - 703.

\bibitem{Liljeros}
{F. Liljeros, C.R. Edling, L.A.N. Amaral, H.G. Stanley and Y. \.{A}berg,
The web of human sexual contacts,
\em Nature,}
{\bf 411} (2001), 907 - 908.

\bibitem{Grisi}
{J.H.H. Grisi-Filho, R. Ossada, F. Ferreira and M. Amaku,
Scale-free networks with the same degree distribution:
different structural properties,
\em Physics Research International,} 
{\bf 2013} (2013), Article ID 234180.

\bibitem{Cohen}
{R. Cohen, S. Havlin, D. ben-Avrahan,
Efficient immunization strategies for computer networks and populations,
\em Physical Review Letters,}
{\bf 91(24)} (2003), 247901.

\bibitem{Newman2005}
{M.E.J. Newman,
Power laws, Pareto distributions and Zipf's law,
\em Contemporary Physics,} 
{\bf 46(5)} (2005), 323 - 351.

\bibitem{Ossada}
{R. Ossada, J. H. H. Grisi-Filho, F. Ferreira and M. Amaku,
Modeling the dynamics of infectious diseases
in different scale-free networks
with the same degree distribution,
\em  Advanced Studies in Theoretical Physics,}
{\bf 7(16)} (2013), 759 - 771.

\end{thebibliography}
\end{document}